\documentstyle[11pt]{article}
\makeatletter
\def\dif{{\rm d}}
\def\deriv{\@ifnextchar[{\@deriv}{\@deriv[]}}
   \def\@deriv[#1]#2#3{\mathchoice%
{{\dif^{#1}#2\over\dif{#3}^{#1}}}{{\dif^{#1}#2/\dif{#3}^{#1}}}%
{{\dif^{#1}#2\over\dif{#3}^{#1}}}{{\dif^{#1}#2/\dif{#3}^{#1}}}}

\def\secteqno{\@addtoreset{equation}{section}%
\def\theequation{\thesection.\arabic{equation}}}
\newcounter{subequation}
\def\thesubequation{\alph{subequation}}
\def\sneqnarray{\stepcounter{equation}\let\@currentlabel=\theequation
\setcounter{subequation}{1}
\def\@eqnnum{{\rm (\theequation.\thesubequation)}}
\global\@eqcnt\z@\tabskip\@centering\let\\=\@eqncr\let\@@eqncr=\@@sne
qncr
$$\halign to \displaywidth\bgroup\@eqnsel\hskip\@centering
 $\displaystyle\tabskip\z@{##}$&\global\@eqcnt\@ne
 \hskip 2\arraycolsep \hfil${##}$\hfil
 &\global\@eqcnt\tw@ \hskip 2\arraycolsep $\displaystyle\tabskip\z@{##}$\hfil
  \tabskip\@centering&\llap{##}\tabskip\z@\cr}
\def\endsneqnarray{\@@sneqncr\egroup $$\global\@ignoretrue}
\def\@@sneqncr{\let\@tempa\relax
   \ifcase\@eqcnt \def\@tempa{& & &}\or \def\@tempa{& &}
   \else \def\@tempa{&}\fi
     \@tempa \if@eqnsw\@eqnnum\stepcounter{subequation}\fi
     \global\@eqnswtrue\global\@eqcnt\z@\cr}
\def\nobiblabels{\def\@lbibitem[##1]##2{\@bibitem{##2}}}
\makeatother
\def\tabaddress#1{{\small\it\begin{tabular}[t]{c}#1\\[1.2ex]\end{tabular}}}

\def\ben{\begin{enumerate}}
\def\een{\end{enumerate}}
\def\beq{\begin{equation}}
\def\eeq{\end{equation}}
\def\bea{\begin{eqnarray}}
\def\eea{\end{eqnarray}}
\def\beann{\begin{eqnarray*}}
\def\eeann{\end{eqnarray*}}
\def\beasn{\begin{sneqnarray}}
\def\eeasn{\end{sneqnarray}}

\def\UBECM{Departament d'Estructura i Constituents de la Mat\`eria\\
   Universitat de Barcelona i Institut de Fisica d'Altes Energies\\
   Av.~Diagonal 647\\
   08028 Barcelona\\
   Catalonia, Spain}

\begin{document}
\thispagestyle{empty}
\title{Faddeev-Jackiw approach to gauge theories and ineffective constraints}
\author{\sc
J. Antonio Garc\'ia\thanks{electronic adress: garcia@rita.ecm.ub.es}\,
and
Josep M. Pons\thanks{electronic adress: pons@ecm.ub.es}
\\
\def\baselinestretch{1.2}
\tabaddress{\UBECM}
}
\date{September 1997}

\def\ppclassif{{\noindent\tt UB-ECM-PF 97/20\\hep-th/9803222\\
 PACS: 04.20.Fy, 11.10.Ef, 11.15.-q}}

\pagestyle{myheadings}
\markright{\sc J.A. Garc\'ia, J.M. Pons,
\rm `F-J and ineffective constraints'}

\catcode`\@=11
\newdimen\ex@
\ex@.2326ex
\def\dddot#1{{\mathop{#1}\limits^{\vbox to-1.4\ex@{\kern-\tw@\ex@
\hbox{\rm...}\vss}}}}
\catcode`\@=\active

\def\baselinestretch{1.2}
\nobiblabels
\secteqno

\maketitle

\begin{abstract}

The general conditions for the applicability of the
Faddeev-Jackiw approach to gauge
theories are studied. When the constraints are
effective a  new proof in the Lagrangian framework of the
equivalence between this method and the Dirac approach is given.
We find, however, that
the two methods may give different descriptions for the reduced phase
space  when ineffective constraints are present. In some cases the
Faddeev-Jackiw approach may lose some constraints or some
equations of motion. We believe that this
inequivalence can be related to the failure of the Dirac conjecture
(that says that the Dirac Hamiltonian can be enlarged to an Extended
Hamiltonian including all first class constraints, without changes in the
dynamics) and we suggest
that when the Dirac conjecture fails the Faddeev-Jackiw approach fails
to give the correct dynamics. Finally we
present some examples that illustrate this inequivalence.
\end{abstract}

\vfill
\ppclassif
\clearpage


\section{Introduction}

Some years ago, Faddeev and Jackiw \cite{FJ} suggested a simple
Lagrangian method for dealing with gauge theories.  The distinctive
feature of this method is the way to eliminate the gauge degrees of
freedom: The reduction of the degrees of freedom to the physical ones is
performed just by direct substitution of the holonomic constraints, derived
from the variational principle, into the Lagrangian.  This substitution
takes place algorithmically, in several steps.  Since the usual
constraints that appear in the tangent space of a gauge theory are not
holonomic (they usually involve velocities), the idea of Faddeev and
Jackiw is to work with the canonical Lagrangian, which takes as a new
configuration space the cotangent space (phase space) of the original
theory.  Then the canonical Lagrangian is at most linear in the
velocities, and all the constraints become holonomic (they are the Hamiltonian
constraints).  Thus, the method of Faddeev and Jackiw avoids the
sometimes cumbersome procedure pioneered by Dirac, and known as the Dirac
method \cite{Dirac}. There is a price to pay nevertheless: the necessity to perform
a non-trivial Darboux transformation at each stage of the new algorithm.

In a recent paper \cite{GP} we have proved, under some general
assumptions, the equivalence of the Faddeev-Jackiw (F-J) method and the
classical Dirac approach.  Here we want
to expand this result by considering the cases where some conditions
required in \cite{GP} for the equivalence proof do not hold.  The proof
in \cite{GP} was produced under some conditions of regularity (summarized
in the first section of that paper).  In particular,  we assumed
that the constraints $\phi_\mu(q,p)$ (primary, secondary...) that appear in the
formalism allow for a canonical representation of the constraint
surface, that is,
there is a change of basis,
\beq
\phi_\mu\to\xi_\mu=M_\mu^\nu(q,p)\phi_\nu, \qquad \det M\not=0,
\label{CRCS}
\eeq
to a new set of functions $\xi_\mu(q,p)$ that represent the same
surface as the original constraints $\phi_\mu(q,p)$, and where the functions
$\xi_\mu(q,p)$ are a subset of a new set of canonical variables.
This assumption, crucial in \cite{GP}, therefore takes for granted that
all the constraints are effective, where by effective constraints we
mean the following: A set of independent constraints is said to be
effective --and ineffective otherwise-- if the one-forms obtained by
differentiating the constraints (that is, their gradients) are all
independent on the constraint surface.  Notice that the dimension of
this space of one-forms is invariant under changes of the type
(\ref{CRCS}).

The canonical transformation associated with (\ref{CRCS}) can not be
realized for an ineffective representation of the constraint surface:
As long as $\det M\not=0$, an ineffective representation will remain so,
and the $\xi_\mu$ will never be a subset of a set of
canonical variables.  Since the assumption that the description of the
constraint surface is effective was made at every stage in the
Faddeev-Jackiw reduction algorithm, this means that only effective
constraints were allowed in our proof in \cite{GP}.  Of course, given an
ineffective representation of the constraint surface, it is always
possible to construct an effective representation for it.  In more
mathematical terms this effective representation is a basis of the ideal
of functions that vanish on the constraint surface.  This effective
representation can be used to characterize this surface geometrically
according to the classification of its constraints as first class and
second class.  This characterization is also given by the rank of the
symplectic structure projected from the phase space to the constraint
surface.

The presence of ineffective constraints introduces some problems in the
general theory of constrained systems.  For instance, one can run into
difficulties with the counting of the true --non gauge-- degrees of
freedom, or with the breakdown of the equivalence between the Dirac
Hamiltonian formalism and the Extended Hamiltonian formalism --where all
first class constraints are included in the Hamiltonian with
independent Lagrange multipliers--, which is nothing but the failure of
the Dirac conjecture \cite{Dirac}.  Besides these theoretical aspects, it is
interesting to analyze this type of constraint because they appear in
some examples that exhibit a rich gauge algebraic structure, examples are
the Siegel model \cite{Siegel} \cite{GR} and some models of
$\cal{W}$-algebras  in Euclidean space \cite{W-alg}.  In this paper we
will not discuss the problems that arise at the quantum level in case of
ineffective constraints. This issue has been recently addressed
for some simple examples from the point of view of the quantum projector
method \cite{Klauder}.

As regards the F-J method, we must remark that if for a given theory
some ineffective
constraints appear in Dirac's stabilization algorithm, then the proof of
equivalence of the F-J method with Dirac's, given in \cite{GP}, does not
hold.  In fact we will see that in some cases this equivalence is broken
because the replacement of ineffective constraints in the variational
principle for the canonical Lagrangian, as it is made in the F-J method
for any constraint, leads to a loss of dynamical information.
In this paper we will show that, in some cases, the presence of
ineffective constraints makes the basic result obtained in \cite{GP}
false, while in other
cases the equivalence still holds.  We will also show that in the cases
when the equivalence does not hold, the correct method is that of Dirac,
because in this case F-J method suffers a loss of dynamical
information while Dirac's method does not.

We will use throughout the paper a notation with finite number of
degrees of freedom, though our results can be generalized to field
theories.
The class of constrained dynamical systems under our consideration will
fulfill the following general properties:

(a) The Hessian matrix
$\partial L /\partial{\dot q}^i \partial{\dot q}^j$ is of constant rank
on the constraint surface.

(b) Throughout the stabilization algorithm,
second class constraints never become first class, or equivalently, the
rank of the Poisson bracket among the constraints cannot decrease during
the stabilization algorithm.

(c) We assume the existence of as many
Noether gauge symmetries as primary first class constraints in
phase space. This assumption goes beyond the results in \cite{ghp}.

(d) The primary constraints are always
effective (this is required by Dirac theory \cite{Dirac}).  In
particular this last point ensures the equivalence between the
Lagrangian approach and the Dirac method \cite{Barcelona} with full
generality.

In this framework we  analyze two questions: a) The reason
for the inequivalence between the Dirac method and the F-J approach
when some ineffective constraints are present. b) The possible
relation between this inequivalence and the failure of the Dirac
conjecture. The first question is addresed in full
generality. As regards the second question, we think it is an
interesting open problem that can
help to understand under which conditions the  Dirac
conjecture fails. One of the essential assumptions in proofs of the Dirac
conjecture, as given for example in
\cite{Henneaux}\cite{cabo}, is that all the constraints are
effective. Indeed, to our knowledge, the counterexamples to this
conjecture always present ineffective constraints at some stage of the
Dirac algorithm. The relation of the Dirac conjecture and ineffective
constraints can be established by an explicit construction of the
gauge generator. In fact, we will see,
by showing some examples, that there are cases where some
secondary first class constraints do not need gauge fixing because they
appear in the gauge generator as ineffective pieces.
Because of a lack of a general theory of
construction of Noether gauge generators for ineffective constrained theories,
we cannot prove this relation in general.

In section 2 we recall the general properties of the F-J method when the
constraints are effective, as
described in \cite{GP}.
In section 3 we analyze the way in which the presence of ineffective
constraints affects the F-J method.  In section 4 we suggest to relate the
failure of F-J method when some ineffective constraints are present to
the structure of the gauge generators and to the Dirac conjecture.
In section 5 we give some examples that illustrate the
incompleteness of the F-J method in such cases. Section 6 is devoted
to conclusions.  In Appendix A we show how the presence of ineffective
constraints is dealt within Dirac's approach.

\section{The Faddeev-Jackiw method for effective constraints}

Let us start by recalling some of the results obtained in our
previous proof in \cite{GP} under the  conditions (a)-(d) of the
previous section. In this
section we add the condition:

(e) All
constraints in the stabilization algorithm are effective.

The F-J reduction procedure \cite{FJ} starts with a general Lagrangian
of the form
\beq
L_c=p_i{\dot q}^i-H_c(q,p)-\lambda^\mu\phi_\mu,
\label{L-org}
\eeq
where $H_c$ is the canonical Hamiltonian, $\lambda^\mu$ are a set of
Lagrange multipliers, and the primary constraints $\phi_\mu$ are taken as
being effective and independent.  By plugging these constraints into the
Lagrangian we can eliminate as many variables as primary constraints,
obtaining the reduced Lagrangian $L$, in terms of a reduced set of
variables that we denote as $x$,\footnote{Notice that the sub or
super-index carries information both on the type of variable --labeled
by the letter-- under consideration and the range of values that it can
take.}
\beq
L=a_s(x){\dot x}^s - H(x)
\label{L'}
\eeq
where $a_s(x)$ are some specific functions that define the Liouville one-form,
and give, after differentiation, the
two-form symplectic structure in the reduced phase space.  We can always perform
a Darboux transformation
\beq
x^s\to Q^r, P_r, Z^a,
\label{Darboux}
\eeq
such that in these new coordinates $L$ takes the canonical form
\beq
L=P_r{\dot Q}^r - H(Q^r,P_r,Z^a).
\label{L-z}
\eeq
The $Z^a$ variables appear only when the two-form defined by $a_s(x)$ is
degenerate, otherwise the functions $a_s(x)$ define the Dirac brackets
in the reduced phase space.  The equations of motion associated to $L$
for the $Z^a$ variables allow for the isolation of a subset of these
variables together with some relations among the original variables
$Q^r, P_r$:
\beq
\frac{\partial H}{\partial Z^a}=0 \Longleftrightarrow Z^{a_1}=f^{a_1}(Q^r, P_r,
Z^{a_2}), \quad f_{a_2}(Q^r,P_r)=0.
\label{za1}
\eeq
Substitution of $Z^{a_1}=f^{a_1}(Q^r, P_r, Z^{a_2})$ into $L$ yields
(see equation (2.9) in \cite{GP})
\beq
L'=P_r{\dot Q}^r - H'(Q^r,P_r) - Z^{a_2}f_{a_2}(Q^r,P_r),
\label{Lred}
\eeq
which again has the structure of the Lagrangian (\ref{L-org}) with
$Z^{a_2}$ as new Lagrange multipliers.
The algorithmic procedure is
thus established and the next step will be the elimination of another set
of variables, by plugging the new constraints $f_{a_2}=0$ into $L'$.

To
summarize, at each stage of the algorithm, we plug the constraints into
the action to get a reduced Lagrangian, and diagonalize (by means of a
Darboux transformation) its associated symplectic form.  If this form is
degenerate we obtain as a byproduct new constraints that can be plugged
again into the reduced Lagrangian.  The procedure continues through a
new diagonalization and ends up when no variables of the type $Z^a$ appear
in the formalism.  At this point we get a non-degenerate symplectic
structure that actually represents the Dirac brackets in the reduced
phase space.

Continuing with the assumption that all constraints are effective the
following results were obtained in \cite{GP}:

(i) The F-J method is
completely equivalent to the Dirac approach for gauge theories.

(ii) The
Darboux transformation (\ref{Darboux}) is the projection on the
constraint surface of a canonical transformation in the whole phase
space that allow for a canonical representation of this surface.

(iii)
The relations $Z^{a_1}- f^{a_1}(Q^r,P_r)=0$ play the role of second
class constraints with respect to a subset of the primary first class
constraints that from now on become second class.

(iv) The $Z^{a_2}$
variables are canonical variables with respect to the remaining set of
primary first class constraints and can be considered as Lagrange
multipliers in (\ref{Lred}).  As it happens with the original Lagrange
multipliers, these $Z^{a_2}$ variables are bound to become either pure
gauge arbitrary variables or functions of the physical variables.  In
any case they will not play any role in the description of the reduced
--physical-- phase space.


\section{Ineffective constraints and the Faddeev-Jackiw method}

The results of the previous section rely on the condition that
all constraints are effective at
each step of the algorithm.  Now let us drop this condition.  At this
point, the constraints in (\ref{za1}) may have appeared in
$\partial H /\partial
Z^a=0$ as ineffective.  Then, the substitution of $Z^{a_1}$ by
$f^{a_1}$ in
(\ref{L-z}) may produce the disappearance of some of the $Z^{a_2}$
variables.  And more: If some of the constraints $f_{a_2}=0$ are
ineffective, it is not legitimate to interpret them as primary
constraints associated to the new Lagrange multipliers $Z^{a_2}$.  All
these circumstances may lead to a loss of dynamical information --in the
form of constraints or equations of motion-- in the F-J method.

In order to deal with these problems we need to develop a new perspective
of the reduction process that does not depend explicitly on the canonical
representation of the constraint surface.  We will use
configuration-velocity space methods and discuss the validity of the F-J
method by studying the Lagrangian equations of motion.  As a byproduct
we will obtain a new proof of the equivalence between the Dirac and F-J
methods when no ineffective constraints appear.

The key idea is to analyze the two processes of reduction implied in the F-J
method, namely the reduction from the Lagrangian (\ref{L-z}) to
(\ref{Lred}) and the reductions of the new constraints $f_{a_2}=0$ in
(\ref{Lred}). The basic difference between the analysis of the present
section and the previous one is  that we now allow
the constraints $Z^{a_1}-f^{a_1}=0$ and $f_{a_2}=0$ to be ineffective.


\subsection{General reduction for holonomic constraints}

Consider a configuration space locally described by the coordinates
$x_a, q_j$.  Suppose a general Lagrangian of the form
$$  L(x_a,q_j;\dot x_a,\dot q_j),
$$
and, for some regular functions $g_a(q)$, let $x_a-g_a(q)=0$ define a
surface in this configuration space.

The pull-back of $L$ to this surface will define the reduced Lagrangian
$L_{red}$ as
$$  L_{red}(q;{\dot q}) = L(g_a(q),q_j;\frac{\partial g_a}{\partial q_i}
{\dot q}_i,{\dot q}_j),
$$
and there is the following relationship between
the Euler-Lagrange derivatives for these two Lagrangians:
\beq
[L_{red}]_{q_j} =
{[L]_{q_j}}\big|_{x=g(q)} +  \frac{\partial g_a}{\partial q_j}
{[L]_{x_a}}\big|_{x=g(q)}.
\label{e-l-relations}
\eeq
In the next subsections we will make use repeatedly of this result.


\subsection{Type 1 problems: elimination of $Z^{a_1}$}

Now let us substitute $Z^{a_1} = f^{a_1}
(Q^r,P_r,Z^{a_2})$ from (\ref{za1}), into (\ref{L-z}).
This defines the partially reduced Lagrangian $L'$,
\beq
L'(Q^r,P_r,{\dot Q}^r, {\dot P}_r; Z^{a_2}) =
L(Q^r, P_r, {\dot Q}^r, {\dot P}_r; f_{a_1}(Q^r,P_r,Z^{a_2}), Z^{a_2}).
\label{L'2}
\eeq
whose equations of motion satisfy, according to (\ref{e-l-relations}),
$$  [L']_{Z^{a_2}} = {[L]_{Z^{a_2}}}\Bigg |_{{Z^{a_1}} =f^{a_1}} +
\frac{\partial f_{a_1}}{\partial Z^{a_2}}
{[L]_{Z^{a_1}}}\Bigg |_{{Z^{a_1}} =f^{a_1}} .
$$
Now, according to (\ref{za1}), and noticing that the
equations of motion for the $Z$ variables are just extremity conditions,
we have
\beq
{f_{a_2}} = 0 \Longrightarrow [L']_{Z^{a_2}} = 0.
\label{implic}
\eeq
The implication is not a two way implication because some of the variables
$Z^{a_2}$ may disappear from $L'$.  This may happen if some of the relations
$Z^{a_1}-f^{a_1}=0$, which are effective constraints, have originally appeared
in (\ref{za1}) in an ineffective form.  Let us produce an example.
Consider $L = P \dot Q + Z_2(Z_1 - f_1(Q,P))^2 + (Z_1 -
f_1(Q,P)) f_2(Q,P)$.  Application of (\ref{za1}) gives $Z_1 -
f_1(Q,P) = 0$ and $f_2(Q,P) = 0$, but when we substitute
$f_1(Q,P)$ for $Z_1$ in $L$ we get $L'= P \dot Q$: The variable $Z_2$
disappears and the constraint $f_2(Q,P) = 0$ is not retrievable from
$L'$.

This analysis implies that we can only guarantee that
$$  [L]_{Z^a} = 0 \Rightarrow  [L']_{Z^{a_2}} = 0 ,
\quad  Z^{a_1} - f^{a_1}(Q, P; Z^{a_2}) = 0.
$$
Denoting the rest of the variables, $Q$ and $P$, as $X$, we have, using
again (\ref{e-l-relations}),
\beq
[L']_X  = {[L]_X}\Bigg |_{{Z^{a_1} = f^{a_1}}}
+ \frac{\partial f_{a_1}}{\partial X}
{[L]_{Z^{a_1}}}\Bigg |_{{Z^{a_1} = f^{a_1}}}.
\eeq
On the surface defined by $f_{a_2} = 0$ the last term vanishes, and
the rest is
\beq
[L']_X\Bigg |_{f_{a_2} = 0}   = {[L]_X}\Bigg |_{[L]_{Z^a}=0}=0.
\eeq
Therefore we arrive at
\beq
\left\{\matrix{[L]_{X} = 0\cr
                    [L]_{Z^a} = 0}\right\} \quad \Longrightarrow \quad
\left\{\matrix{[L']_{X}=0\cr
 [L']_{Z^{a_2}} = 0\cr
Z^{a_1} - f^{a_1}(Q,P; Z^{a_2})=0}\right\}.
\label{implic2}
\eeq
This one-way-only implication is the type 1 problem with the F-J method. The
equivalence only
holds when (\ref{implic}) is indeed an equivalence, that is, when
$${f_{a_2}} = 0 \Longleftrightarrow [L']_{Z^{a_2}} = 0.$$
This equivalence
is guaranteed if the $Z^{a_1}$ type variables
are auxiliary variables
\footnote{Auxiliary variables are a set of variables that can
be obtained (as a set) by using their own equations of motion in terms of
the rest of the variables that describe the system.  For details see
\cite{Henneaux}.} {\it i.e.}, $[L]_{Z^{a_1}}=0 \Leftrightarrow
Z^{a_1}-f^{a_1}=0 $.

We conclude that there is a possible loss of dynamical information when
the original Lagrangian $L$ is partially reduced to $L'$ by plugging
into it the relations ${Z^{a_1}} = f^{a_1}(P, Q; Z^{a_2})$.  This loss
of information originates in the one-way implication displayed in
equation (\ref{implic}). This non-equivalence has its roots
in the fact that some of these relations may appear within ineffective
constraints.


\subsection{Type 2 problems: reduction to the surface $f_{a_2}(Q,P) =
0$}

There is a second source of problems, also related to ineffectiveness,
that haunts the F-J method.  Suppose we still have an equivalence in
(\ref{implic2}) and let us complete the reduction of the
Lagrangian (\ref{L-z}) by plugging the constraints $f_{a_2}(Q,P) = 0$ into
(\ref{Lred}).  Here we consider that all these constraints are
independent.  In case they are not, their number will be reduced
accordingly and so will be the number of $Z^{a_2}$ variables appearing
in $L'$.  As we already note in (\ref{Lred}), $L'$ takes the form
\bea
\nonumber
L'(Q^r, P_r, {\dot Q}^r;{\dot P}_r, Z^{a_2}) &=&
P_r \dot Q^r -  H'(Q^r,P_r) - Z^{a_2} f_{a_2}(Q^r,P_r)\\
&=:& A(Q^r,P_r,{\dot Q}^r, {\dot P}_r) -
Z^{a_2} f_{a_2}(Q^r,P_r),
\label{L'-z}
\eea
that is, $L'$ is at most linear in the variables $Z^{a_2}$. Let us change
the variables $Q,P$ to variables $y_m,x_{a_2}$ such that\footnote{The
number of $x$-type variables may be larger than the number of functions
$f_{a_2}$ because one of these functions being ineffective may kill more
than one degree of freedom. This is the case for instance of the square
of the norm of a vector in Euclidean space. For the sake of simplicity
we use the same indices for the $x$ variables and the $f_{a_2}$ functions.}
\beq
f_{a_2}(Q,P) = 0 \Longleftrightarrow x_{a_2} = 0.
\label{fa2}
\eeq
Notice that the equivalence (\ref{fa2}), does not guarantee the
effectiveness of  $f_{a_2}(Q,P)$; for instance it could be that
$f_{a_2}(Q,P) = ({x_{a_2}})^2$. Now
consider the further reduction of $L'$,
to the surface $x_{a_2} = 0$:
\beq
L_{R} (y) = L'(x_{a_2}=0, y_m,
{\dot x}_{a_2} =0, {\dot y}_m).
\label{LR}
\eeq
Notice that $Z^{a_2}$ disappears from $L_{R}$. Applying
(\ref{e-l-relations}), we have
$$[L_{R}]_{y_m} = {[L']_{y_m}}\Bigg |_{x_{a_2}=0}.
\label{l-red}
$$
Let us now discuss separately the two cases we can find according to the
effectiveness or ineffectiveness of the constraints.

A)  Consider the case when the constraints
$f_{a_2}(P,Q) =0 $ are truly effective: Without loss of generality
we can take the variables $x,y$ such that
$f_{a_2}(P,Q) = x_{a_2}$.
In this case,
$$[L_{R}]_{y_m}={[L']_{y_m}}|_{x_{a_2}=0} =
{[A]_{y_m}}|_{x_{a_2}=0}.
\label{l-red1}
$$
On the other hand,
$${[L']_{x_{a_2}}}\big|_{x_{a_2}=0} =
{[A]_{x_{a_2}}}\big|_{x_{a_2}=0} - Z^{a_2}.
\label{l-red2}
$$
So we have,
\beq
\left\{\matrix{ [L']_{Z^{a_2}}=0\cr
[L']_{y_m}=0\cr
[L']_{x_{a_2}}=0}\right\}\quad\Longleftrightarrow\quad
\left\{\matrix{
x_{a_2}=0\cr
[L_R]_{y_m}=0\cr
Z^{a_2}=[A]_{x_{a_2}}\big|_{x=0}}\right\}.
\label{efectiu}
\eeq

As it is argued in \cite{GP}, the $Z^{a_2}$ variables are irrelevant
because either they are gauge variables or they become determined
through constraints as functions of the physical (gauge invariant)
variables.  In any case we can get rid of them and hence the equations
\beq
Z^{a_2}=[A]_{x_{a_2}}\Big|_{x_{a_2}=0}
\label{complete}
\eeq
can be ignored.
The equations for the relevant set of $y$ variables (until subsequent
reductions further cut down this set) are therefore $[L_R]_{y_m}=0$.

This proves the correctness of this stage of the F-J reduction procedure
as long as ineffective constraints do not appear in the formalism.
If at each stage, no ineffective constraints appear, we have produced
a new proof of the correctness of the F-J method, that is, its
equivalent to the canonical Lagrangian analysis which in turn is
equivalent to the Dirac's method.

B) Consider, for the sake of simplicity, that all the constraints
$f_{a_2}(Q,P)=0$ are ineffective.
In
such case (\ref{efectiu}) is modified to
\beq
\left\{\matrix{[L']_{Z^{a_2}}=0\cr
[L']_{y_m}=0\cr
[L']_{x_{a_2}}=0}\right\}\quad\Longleftrightarrow\quad
\left\{\matrix{x_{a_2}=0\cr
[L_R]_{y_m}=0\cr
[A]_{x_{a_2}}\big|_{x_{a_2}=0}=0}\right\}.
\label{inefectiu}
\eeq

Notice that the reduced equations of motion, $[L_R]_{y_m}=0$, are
potentially incomplete because of the presence of new equations for the
$y$ variables: those given by

\beq
[A]_{x_{a_2}}\Big |_{x_{a_2}=0}=0.
\label{incomplete}
\eeq

This is the type 2 problem with F-J reduction method. Only when these
new equations are empty or do not add new information to
the reduced equations derived from $L_R$ (as in the case where at some
stage of the F-J algorithm the set $f_{a_2}$ is empty), can we say that
the Faddeev--Jackiw method still works.  Otherwise, there is a loss of
dynamical information, for the equations of motion derived from the
reduced Lagrangian $L_R$ are not the whole set of equations of motion
for the reduced variables. Whether this loss of information consists in
the loss of some constraints or of true equations of motion (that
is, equations with velocities in the lhs) will be explored in the next
subsection.

In a general case, when some of the constraints $f_{a_2} = 0$ are
effective and some are ineffective, both types of equations,
(\ref{complete}) and (\ref{incomplete}), will appear, the first
associated with the effective constraints and the second with the
ineffective ones.  The potential incompleteness of the F-J method comes
in this case from this last type of equations.


\subsection{Losing constraints and equations of motion in the type 2
problems}

The previous results can be reformulated in a more transparent way by
using a canonical representation for the  variables describing
the constraint surface.  In order to simplify the notations let us
suppose that all the constraints $f_{a_2}$ are ineffective and recall
(\ref{fa2}),
$$f_{a_2}(Q,P) = 0 \Longleftrightarrow x_{a_2} = 0.$$
In
order to cover the most general case, take for the coordinates $x_{a_2},
y_m$ the canonical form $x_{a_2}=\{Q^s,P_s,P_u\}$ and $y_m=\{Q^u, Q^t,
P_t\}$ where coordinates and momenta with the same label are canonical
pairs.  In these new coordinates the Lagrangian (\ref{L'-z}) can be
written as
\beq
L'=P_u{\dot Q}^u+ P_s{\dot Q}^s+ P_t{\dot Q}^t -
H'(Q^u,P_u,Q^s,P_s,Q^t,P_t) - Z^{a_2}{f}_{a_2}(Q^s, P_s, P_u; Q^u,
Q^t,P_t),
\eeq
where we have kept, for simplicity, the notations of (\ref{Lred}) for all
functions involved.  Now
$$f_{a_2} = 0 \Longleftrightarrow Q^s = P_s = P_u = 0.$$
The reduced Lagrangian (\ref{LR}) becomes
\beq
L_R=P_t{\dot Q}^t-H_R(Q^u,Q^t,P_t),
\eeq
where $H_R=H'(Q^u,P_u=0,Q^s=0,P_s=0,Q^t,P_t)$.
As we know from the previous analysis this Lagrangian may not contain
all the dynamical information of the reduced dynamics. Its equations of
motion $[L_R]_{y_m}= 0$ are
\beq
{\dot Q}^t-\frac{\partial H_R}{\partial P_t}=0, \quad
-{\dot P}_t-\frac{\partial H_R}{\partial Q^t}=0, \quad \frac{\partial
H_R}{\partial Q^u}=0,
\eeq
where the last set of equations are constraints.  The dynamical
information loss is contained in the equations of motion
(\ref{incomplete}) $[A]_{x_{a_2}}\big|_{x_{a_2}=0}$. Thanks
to the canonical representation of the constraint surface, these
equations take the simple form
\beq
\frac{\partial H'}{\partial P_s}\Big|_{x_{a_2}=0}=0, \quad
-\frac{\partial H'}{\partial Q^s}\Big|_{x_{a_2}=0}=0, \quad
{\dot Q}^u-\frac{\partial H'}{\partial P_u}\Big|_{x_{a_2}=0}=0.
\label{versus}
\eeq

Notice that the first two sets of equations in (\ref{versus}) are just
constraints, maybe new ones, maybe not, whereas the last one contains
only true equations of motion, which are all new at this level.  This
difference
--possible new constraints versus true equations of motion-- has its
roots in the first and second class character of the surface $f_{a_2} =0$,
which is revealed after the effectivization of its defining constraints.
Part of these effectivized constraints, $P_u=0$ , are first class and
the rest, $Q^s=0,P_s=0$, second class.  Then, as we see in
(\ref{versus}), the possible loss of constraints comes from the sector
of the second class effectivized constraints,
whereas the loss of equations of motion comes from the
sector of the first class effectivized constraints.

Summing up: There are two sources, both related to ineffective
constraints, for incompleteness in the F-J
method: the reduction of the $Z^{a_1}$ type variables and the reduction
of the ineffective constraints among the set of functions $f_{a_2}$.
Sometimes this incompleteness amounts to a loss of constraints, whereas
in some other cases there is a loss of equations of motion.  If the
basis of the ideal of functions vanishing on the surface defined by the
ineffective functions $f_{a_2}=0$ contains first class constraints, then
we conclude that there is a true loss of dynamical information in the
form of equations of motion, {\it i.e.,} involving velocities.  Whether
there is a real loss of information for the physical variables or not,
cannot be decided until the F-J algorithm is completed.  Our examples in
section 5 will show cases where this loss is real.


\section{Relation with Dirac's conjecture}

In this section we want to examine the failure of F-J reduction method
in presence of ineffective constraints from another perspective.  Here
we will work in Dirac's formalism in order to consider the gauge
generators of the theory and how the process of gauge fixing is
modified by the presence of ineffective constraints.

Constraint analysis in Dirac formalism has at least two conceptually
different applications: the stabilization algorithm, on one side, and
the construction of the gauge generators, on the other.

The gauge generators are made up of first class constraints in a chain
that involves some arbitrary functions and their derivatives.  Since we
are working with the canonical Lagrangian, there is a role, too, for the
Lagrange multipliers as new variables.  In a given theory, the complete
gauge generator contains as many arbitrary functions as there are
independent gauge transformations in the theory.  This number
coincides with the number of primary first class constraints.

Notice that these two aspects of the constraint analysis become
complementary with regard to the determination of the correct number of
(physical) degrees of freedom of our theory.  The determination of the
constraint surface is a first step in this direction.  Next we
need to know the gauge generator of the theory in order to
eliminate further the spurious degrees of freedom associated with the gauge
symmetries.

All the constraints appearing in $G$ are first class (this fact
was first proved by Dirac) \cite{Dirac}, but nothing prevents that some
of them be ineffective.  Even more, there are examples \cite{GR} where
the effectivization of some constraints involved in $G$ is second class!

The presence of ineffective constraints at the secondary,
or tertiary, etc., level of
the constraint algorithm is intrinsic to the dynamical system under
consideration.  As far as the stabilization algorithm is concerned we
can, at each step, make these ineffective constraints effective by a wise
choice of a new set of functions that generate the ideal of functions
vanishing at the constraint surface that has been determined so far.  In
this sense the stabilization algorithm is essentially unaffected by the
presence of ineffective constraints.  Certainly, this ``effectivization''
of constraints is the standard way to proceed in Dirac's method
\cite{Barcelona} \cite{Gotay} but it is not mandatory.  In fact, in
appendix A we sketch how the stabilization algorithm works in the presence
of ineffective constraints.

With regard to the construction of the gauge generator, this
``effectivization'' of constraints is not allowed\footnote{Notice that every
ineffective
constraint is first class, regardless
of whether its ``effectivization'' is first or second class.}.
Let us elaborate on
this important point. As we show in \cite{GP2} a generator $G$ of a
Noether symmetry depending on $q, p, t, \lambda, \dot\lambda...$,
where $\lambda$ denotes the set of Lagrange multipliers associated with
the primary constraints, is characterized by the property
\beq
\frac{DG}{Dt}+\{G,H_D\}=pc, \quad
\frac{D}{Dt}=\frac{\partial}{\partial t}+
\dot\lambda \frac{\partial}{\partial\lambda}+
\ddot\lambda \frac{\partial}{\partial\dot\lambda}+... \ ,
\label{gauge}
\eeq
where $H_D=H_c+\lambda^{m_1}\phi_{m_1}$, $H_c$ is the canonical
Hamiltonian and $\phi_{m_1}$ are the primary constraints, and $pc$
stands for a linear combination of primary constraints. The Noether
transformations for the canonical Lagrangian $L_c = P_i\dot Q^i - H_D$
are defined by
$$\delta Q = \{Q,\,G\}, \quad \delta P = \{P,\,G\}
$$
and $\delta \lambda^\mu$ is defined so that $\delta L_c$ becomes a total
time derivative. In some cases
a pure gauge
generator depending only on $q,p,t$ may be constructed and the
condition (\ref{gauge}) splits into (see also \cite{BGGP})
\beq
\frac{\partial G}{\partial t} + \{G,\,H_c\} = pc, \qquad
\{G,\,pc \} = pc.
\label{g-noet}
\eeq
These last conditions are, in general, more restrictive, and as a
consequence a solution $G$ of (\ref{g-noet}) may not exist while a
solution $G$ to (\ref{gauge}) can still be constructed.
With the notation $\phi_{m_k}$ for the $k$-ary (primary, secondary...)
first class constraints we can write the
gauge generator in the form
\beq
G=\sum \mu^{m_k}(q,p,t,\lambda,\dot\lambda,...)\phi_{m_k}(q,p)
\eeq
and  recover the general results of \cite{Zanelli} under weaker
assumptions. For details see \cite{GP2}.

As is well established, \cite{Zanelli}, \cite{Henneaux},
\cite{Barcelona},
the standard counting of the
degrees of freedom for systems that do not exhibit ineffective
constraints is as
follows: If the dimension of the original phase space is $2N$,  the
number of first class constraints is $m$, and the corresponding (even)
number of second class constraints is $2s$, the total number of degrees
of freedom is $2F=2N-2m - 2s$.

If there are ineffective constraints in the gauge generators, then we
must not introduce any gauge fixing constraints for them.  This is so because
these constraints do not generate any motion in the constraint surface
and therefore they do not transform the dynamical trajectories.  This
means that the gauge fixing constraints must only be included
for the secondary, tertiary,... first
class effective constraints in $G$. For details on the gauge fixing
procedure, see \cite{Shepley}.  The counting of degrees of freedom
is obviously affected by this circumstance, and for instance, the final
number of physical degrees of freedom may be odd, as it happens in some
of the examples in the next section.

Let us now make contact with Dirac's conjecture.  In a modern
interpretation, this conjecture says that it is always possible to
enlarge the Dirac Hamiltonian (which already contains the primary first
class constraints with their Lagrange multipliers) with the addition of
all the remaining, secondary, tertiary, etc. first class constraints,
and their new associated Lagrange multipliers, without any change of the
dynamics of the theory and its physical interpretation.  This enlarged
Hamiltonian is known as the Extended Hamiltonian.  It can be proved that
if all constraints are effective, the Extended Hamiltonian and the Dirac
Hamiltonian give equivalent results, and the reason is that we can
always introduce a gauge fixing for both Hamiltonians that yield the same
dynamics
in the same reduced phase space \cite{Henneaux}.  The differences arise
when there exist ineffective constraints.  In such case, if the first
class constraints that are added to the Dirac Hamiltonian to define the
Extended one, are given in an effective representation, then there
will be gauge fixings for all these first class constraints, even for the
ones that
come from the effectivization of ineffective ones.  So we see that in
this case there is an ``excess of gauge fixing'' that makes the Extended
theory inequivalent to the one described by the original Dirac
Hamiltonian. We think it plausible, and all  examples that we have
studied support it, that there is a link between the failure of the
Dirac conjecture and
the failure of the F-J method to describe the correct reduced
dynamics in such a way that
\beq
\mbox{ Dirac conjecture fails} \Longrightarrow \mbox{F-J method fails},
\eeq
but we are not able to prove this relation because of a lack
of a general theory
of the construction of gauge generators for ineffective constrained
theories.

As a final comment let us mention that,
using the antibracket cohomology \cite{BBH} our results  can be
reformulated  as follows:  In the case of effective constraints
it is proved in \cite{Henneaux} that the BRST cohomology at ghost number
zero consist of all the observables of a given physical theory.  On the
other hand, the BRST cohomology is invariant with respect to the
elimination of the auxiliary variables
(in our notations $Z^{a_1}$ in the first step of
reduction and  $Z^{a_2}, x_{a_2}$ in the second step) \cite{BBH} and
therefore we can conclude that the reduction process produces the same
results, that is, the reduced (by eliminating the auxiliary
variables)  theory
is completely equivalent with the original theory.  But in the case of
ineffective constraints these theorems no longer apply because some of
the $Z^{a_1}, Z^{a_2}$ type variables are not auxiliary variables.
The loss of
dynamical information produced in the reduction process by this fact
was analyzed in section 3. It will be of interest, specially for
field theories, to
analyze this loss of dynamical information from a cohomological
perspective. In this respect, it may be helpful to analyze how the
symmetries of the theory are altered in the reduction process.


\section{Examples}

In order to exhibit in a transparent way the inequivalence results
obtained in section 3 and their relation with the failure of the Dirac
conjecture, we choose some
simple examples. First, we choose a
model that by some ineffective $Z^{a_1}$-type variables lose a
first class constraint
upon reduction. By constructing the gauge generator we show that the
Dirac conjecture is violated. Then we analyze another model that
presents two secondary ineffective constraints (effectivized second
class). The model is such that the equations (\ref{incomplete})
give a new constraint. Only when this new constraint is not
considered in the model does the F-J
approach give the correct reduced dynamics, because then equations
(\ref{incomplete}) are empty. By an explicit
construction of the gauge generator we show that the Dirac  conjecture is
violated. The model can be extended (by adding a new secondary
constraint) in such a way the the Dirac conjecture is now valid but
still the F-J approach fails. The next example contains one
ineffective constraint that upon effectivization is first class. Then
we show that the F-J approach lose an equation of motion. By an
explicit construction of the gauge generator we show that the system
violates the Dirac conjecture
\footnote{In this section we use only subindex notation for clarity in
exposition.}.


\subsection{Type 1 problems}

In order to exhibit a simple case where the substitution of the
$Z^{a_1}$ variables (the first step in the F-J reduction) produces a
loss of information,
let us consider the canonical Lagrangian
\beq
L_c=\sum_{i=1}^{4}{\dot q}_ip_i-
\sum_{i=1}^4\frac12 p_i^2-p_3q_2-q_1q^2_2-\lambda_1 p_1 -\lambda_2 p_2.
\eeq
Here $\lambda_1,\lambda_2$ are Lagrange multipliers.

In Dirac's constraint analysis, the Dirac
Hamiltonian is
\beq
H_D= \sum_{i=1}^4\frac12 p_i^2
 +p_3q_2+q_1q^2_2+\lambda_1 p_1 +\lambda_2 p_2.
\eeq
The stabilization of the primary constraints $p_1=0$ and $p_2=0$ gives
$q_2^2=0$ and
$p_3=0$ as secondary constraints. A subsequent stabilization determines
$\lambda_2=0$. The only Lagrange multiplier
that remains arbitrary is $\lambda_1$. Reducing the second class
constraints and introducing a gauge fixing for the constraint $p_1=0$
(for instance, $q_1=0$), we find that
the final reduced dynamics for the physical degrees of freedom is given
by the equations of motion
\beq
{\dot p_4}=0, \quad {\dot q}_4=p_4, \quad {\dot q}_3=0.
\eeq
Notice that the total number of degrees of freedom is odd, i.e.,
$q_3(0), q_4(0), p_4(0)$.

The gauge generator for this theory is
\beq
G=\epsilon\big(q_2p_3+q_1q_2^2+\lambda_1p_1+\lambda_2p_2\big)+
\dot\epsilon\big(q_2p_2-q_1p_1\big).
\eeq
>From the structure of this generator we observe that:

(a) The number of arbitrary parameters in $G$ is equal to the number of
primary first class constraints.

(b) Only the piece containing $p_1$
generates true gauge
transformations on the constraint surface, all other terms are
ineffective (in fact, under the Lagrangian equations of motion,
$\lambda_2$ is the time derivative $\dot q_2$ of the constraint $q_2$).

(c) The secondary first class constraint $p_3=0$ does not
generate a gauge transformation on the constraint surface, that is, the Dirac
conjecture is violated.

Because the Dirac conjecture fails the Extended Hamiltonian formalism also
fails.
To see this let us construct the Extended Hamiltonian
\beq
H_E= \sum_{i=1}^4\frac12 p_i^2+p_3q_2+q_1q^2_2+\lambda_1 p_1 +\lambda_2 p_2+
\lambda_3 q_2 +\lambda_4 p_3.
\eeq
Upon eliminating the second class constraints $q_2, p_2$ and using
the corresponding (trivial) Dirac bracket we obtain
\beq
H'_E= \frac12 p_4^2+  \frac12
p_3^2+\lambda_1 p_1 +\lambda_4 p_3.
\eeq

Now to fix the dynamics we need two gauge fixing conditions.
Take for example $q_1=0, q_3=0$; the result is
free particle motion in the space $(q_4,p_4)$. This result is
different from the correct Dirac analysis and also different from the
one given by the F-J method as we will see.

Applying the F-J
method we eliminate the primary constraints $p_1,p_2$ to obtain
\beq
L'_c={\dot q}_3p_3+{\dot q}_4p_4 -\frac12 p_4^2-\frac12
p_3^2-p_3q_2-q_1q^2_2,
\eeq
and the variables $q_1,q_2$ are now $Z$-type variables. The
equations of motion associated with these variables are $q_2^2=0$
and $p_3+2q_1q_2=0$ respectively. The first relation allows for the
isolation of the variable $q_2$ as $q_2=0$, which is a $Z^{a_1}$-type
variable. The second one implies $p_3=0$, which is a constraint of
the type $f_{a_2}$. The F-J reduction procedure first dictates to
plug $q_2=0$ into the canonical
Lagrangian. We get
\beq
L_R= {\dot q}_3p_3+ {\dot q}_4p_4 -\frac12 p_3^2 - \frac12 p_4^2.
\eeq

This Lagrangian has no dependence whatsoever on $q_1$, and as a
consequence the constraint $p_3=0$ can not be obtained from
$L_R$. As expected, the reason for this
information loss is the ineffective character
of the constraint $q_2^2=0$.
The reduced Lagrangian $L_R$ is regular and no new constraints arise
from its dynamics.
We can conclude that the violation of the
Dirac conjecture is related to the failure of the F-J reduction
process to give the correct reduced dynamics.


\subsection{Type 2 problems: some constraints are missing}

In this example we analyze a system with two primary
first class constraints that have a tertiary constraint in the Dirac
method that is missing in the F-J approach.  Here we want to illustrate
the
failure of the F-J method when the constraints (\ref{fa2}) contain a set
of second class effective constraints among themselves at some level of
the reduction process.
Consider the following Lagrangian:
\beq
L_c=\sum_{i=1}^n{\dot q}_ip_i -  H_c(q,p)-\lambda_1 p_1 -\lambda_2 p_2,
\eeq
where
\beq
 H_c(q,p) := H_R(q_r,p_r)+ q_1 p_3^2 + q_2 q_3^2 +
F(q_r,p_r)p_3,\quad i=1....n,\quad r\ge4.
\eeq
$H_R$ and $F$ are functions that we do not need to specify. The variables
$q_1,q_2$ are considered, by construction, as $Z^{a_2}$
type variables.

The application of the Dirac method is straightforward: The
two primary first class constraints $p_1=0, p_2=0$ produce the
ineffective constraints $p_3^2=0, q_3^2=0$ which define a pair of second
class constraints upon effectivization, namely $p_3=0, q_3=0$.  A
new stabilization of these constraints yields $F(q_r,p_r)=0$.
This new constraint can give rise to other new constraints upon
the application of the stabilization algorithm. Suppose for
definiteness that there are no new constraints, that is,
\beq
\{F,H_R\}=0.
\eeq
In that case the gauge generator is
\bea
\nonumber
G&=&\Big(\frac{q_1}{q_2}\dddot{\epsilon}_2 -\frac{1}{4q_2}
{\dddot\epsilon}_1+\frac{3\lambda_1}{2q_2}{\ddot\epsilon}_2
+\frac{{\dot\lambda}_1}{2q_2}{\dot\epsilon}_2\\ \nonumber
&+&q_1{\dot\epsilon}_1+\lambda_1\epsilon_1+2q_1(2q_1{\dot\epsilon}_2-
{\dot\epsilon}_1)+\frac{\lambda_2}{4q_2^2}(-4q_1{\ddot\epsilon}_2+
{\ddot\epsilon}_1-2\lambda_1{\dot\epsilon}_2)\Big)p_1\\\nonumber
&+&\Big(\frac12{\dddot\epsilon}_2+3q_2{\dot\epsilon}_1+\lambda_2\epsilon_1
-4q_2q_1{\dot\epsilon}_2\Big)p_2\\\nonumber
&+&\Big(\frac{q_1}{q_2}{\ddot\epsilon}_2-\frac{1}{4q_2}{\ddot\epsilon}_1
+\frac{\lambda_1}{2q_2}{\dot\epsilon}_2+q_1\epsilon_1\Big)p_3^2\\\nonumber
&+& \Big(\frac12{\ddot\epsilon}_2+q_2\epsilon_1\Big)q_3^2+
(2q_1{\dot\epsilon}_2-{\dot\epsilon_1})q_3p_3\\\nonumber
&-&{\dot\epsilon}_2Fq_3+\epsilon_2F^2+\epsilon_1Fp_3.
\eea
This gauge generator has an effective action only on the primary first
class constraints. The rest of the action is ineffective and as a
consequence does not produce gauge transformations on the constraint
surface. This means that the only gauge fixings needed are for the
primary first class constraints $p_1=0,p_2=0$.  Then we
conclude that the first class constraint $F=0$ does not need any gauge
fixing. The system violates the Dirac conjecture and the extended
formalism fails.

On the other hand, by noting that, after elimination of the
constraints $p_1=0$ and $p_2=0$, the variables $q_1$ and $q_2$
are $Z$-type (see (\ref{Lred})), the F-J method produces
the reduced Lagrangian
\beq
L_R=p_r{\dot q}_r-H_R(q_r,p_r).
\eeq
There are no new constraints and the algorithm stops here. It is clear
that the constraint that appear in the Dirac formalism, namely
\beq
F(q_r,p_r)=0,
\eeq
is not present in the F-J approach.  Note that this constraint come from
the equations of motion (\ref{incomplete}) that the F-J approach is not
able to produce.  As we expected from the general analysis given in the
previous sections, the two procedures yield very different outcomes for
the description of reduced phase space.  The two approaches coincide
only if $F=0$ because the equations (\ref{incomplete}) are in this case
empty.


\subsection{Type 2 problems: some equations of motion are missing}

This example is designed to illustrate the failure of the F-J
method in the case when the constraints (\ref{fa2}) contain, upon
effectivization, only first class constraints.
It illustrates also the case
when the reduced phase space may have an odd number of degrees of
freedom.  Consider the Lagrangian
\beq
L_c=\sum_{i=1}^np_i{\dot q}_i- H_c(q,p)-\lambda p_2,
\label{ejFC}
\eeq
where
\beq
H_c(q,p) := H_R(q_r,p_r) +q_2 p_1^2+F(q_1,q_r, p_r) p_1, \quad
i=1...n,\quad r\ge 3,
\eeq
where $F$ and $H_R$ are functions that we do not need to specify.

In the Dirac analysis, the theory contains a primary first class
constraint $p_2=0$, which upon stabilization gives rise to a new
ineffective constraint $p_1^2=0$.
The gauge generator
\beq
G=(\epsilon-2\epsilon\frac{\partial F}{\partial q_1}) p_2 +\dot\epsilon p_1^2
\eeq
contains an ineffective piece that does not need a
gauge fixing. The model violates the Dirac conjecture and  presents a
weakly (that is, only on shell) gauge
invariant observable $q_1$.  As we expected from the general analysis, the
F-J procedure is unable to reproduce this result because it
considers only the strong (on and off shell) gauge invariant observables.
The equations of motion that
will be lost in F-J method, (\ref{incomplete}), are
\beq
{\dot q}_1 - F(q_1,q_r, p_r)=0.
\label{[a]}
\eeq
The system presents a phase space with an odd number of degrees of
freedom.

Now let us analyze the reduced dynamics that result from the F-J
approach.  A direct application of the F-J ideas gives rise to the
reduced Lagrangian
\beq
L_R={\dot q}_rp_r-H_R(q_r,p_r).
\eeq
As expected the F-J analysis lose the equations
(\ref{[a]}).  We know from our general
analysis that the equation for $q_2$ can be eliminated via a gauge
fixing procedure, i.e., the coordinate $q_2$ is a pure gauge variable.



\section{Conclusions}

In this paper we have studied in detail the F-J method
for gauge theories in the presence of ineffective constraints.
We have singled out the two different sources (type 1 and type 2
problems) of incompleteness that the F-J reduction method may suffer in
this case.  As a byproduct of this analysis we obtained a new proof of
the equivalence between the F-J reduction algorithm and that of Dirac
when ineffective constraints are not present.  This new proof is based
on Lagrangian methods.

The type 1 problems produce the loss of some constraints.  The type 2
problems may be split into two cases: the possible loss of constraints
and the loss of equations of motion.  Our analysis allows one to
identify when the type 2 problems will lead to the first or the second
case.  The loss of constraints is associated with sets of ineffective
constraints of the type $f_{a_2}$ (\ref{za1}), that become second class
upon effectivization, whereas the loss of equations of motion is
associated with first class constraints.

We give examples of every situation and we show in a specific case
(Section 5.1) that the Dirac method, the F-J approach, and the
Extended
Dirac method may give three different dynamics. The correct one is, of
course, the Dirac dynamics, which is always equivalent to the Lagrangian
formulation.

The structure of the gauge generators and the gauge fixing procedure
turns out to be one of the keys for an understanding of the source of
problems originated by the presence of ineffective constraints.  The
discussion of these issues is neatly illustrated in the examples.
The example in Section 5.3  has an odd number of degrees of freedom,
completely compatible with a canonical formulation.  The presence of
ineffective constraints in the gauge generator makes the gauge fixing
procedure different from the standard case, for ineffective constraints
do not require any gauge fixing.  This is exactly the cause of failure
of the Dirac conjecture when the Extended Hamiltonian is built by using
the effectivized form of the constraints.  We then suggest
that the failure of Dirac's conjecture always implies the failure, in
the sense of incompleteness, of the F-J method.

Finally, a word is in order concerning the treatment of ineffective
constraints in the framework of the stabilization algorithm for the
standard Dirac  procedure.  We have devoted appendix A to showing that
the stabilization algorithm works perfectly well in the presence of
ineffective constraints. The only difference is that the new
generation of constraints, as descendants of the ineffective ones,
do not appear at the next level of stabilization but at
second, or higher order, depending upon the degree of
effectiveness of the constraints involved. These new constraints,
as stabilization conditions of
ineffective constraints, are also ineffective. But as concerning the
stabilization algorithm, one can work equally well with the
corresponding effectivized constraints.


\section*{Apendix A: Ineffective constraints and Dirac's method}

In this appendix we show that the Dirac method can be applied when
some constraints are ineffective. We can choose either to stabilize
the ineffective constraints or the effectivized ones.
Consider the simplest case we can think of an ineffective
constraint. Let $f = \phi^2$ be such a constraint, which represents the
same surface --the points where $f$ vanishes-- as $\phi$.  We
represent this surface by the notation
$$f \simeq 0,
$$
or, equivalently
$$\phi \simeq 0.
$$
Here we can be working either in the tangent space or in the cotangent
space (phase space) of a constrained dynamical system.  Let us assume
also that $\phi$ is indeed effective.  Let us suppose that we are performing a
stabilization algorithm for some dynamics defined by the vector field
$X$. Notice that
$$X(f) = 2 \phi X(\phi) \simeq 0,
$$
so one may be tempted to claim that the stabilization algorithm has
finished, for the action of $X$ on $f$ vanishes in the constraint
surface. But this is incorrect, as one can check by going to the next
order. In fact, the requirement
$$X^2(f) = 2 (X(\phi))^2 + 2 \phi X^2(\phi) \simeq 0,
$$
enforces the new constraint
$$(X(\phi))^2 \simeq 0,
$$
which is again ineffective but defines the same surface as
$$X(\phi) \simeq 0.
$$
Thus we notice that, in this case, the stabilization of a ineffective
constraint does not stops at first order, even though the first order
stabilization
does not introduce new constraints. Moreover, we eventually get
the same new restriction, that is, $X(\phi) \simeq 0$, as the
stabilization of an
effective constraint, $\phi$, gives at first order.

To understand why
it is so, we should consider the meaning of the stabilization algorithm.
The dynamics generated by the vector field $X$ defines trajectories
$$x(t) = e^{tX} x,
$$
where $x(0) = x$.  Tangency of these trajectories to the surface defined
by $f = 0$ means that we must require $f(x(t)) = 0$ for any $x = x(0)$
such that $f(x) = 0$. But
$$ f(x(t)) = (e^{tX} f)(x) = f(x) + t (Xf)(x) + \frac{1}{2} t^2
(X^2f)(x) + ... \,,
$$
and we get an infinite set of requirements,
$$(Xf)(x) = 0, \quad (X^2f)(x) = 0,\,... \ .
$$
In general only a few of these terms will introduce new restrictions.
Notice that when  $f$ is ineffective, then the fact that the
first order requirement, $(Xf)(x) = 0$, is automatically satisfied does
not imply that the second order, $(X^2f)(x) = 0$, is satisfied.

In conclusion, when dealing with
ineffective constraints the stabilization algorithm does not finish at
the level where we find no new restrictions. We must proceed further
until we are sure that all the tangency conditions have emerged.

According to these remarks, there are two ways to deal with ineffective
constraints within the framework of Dirac's method.  Either we proceed
through the lines sketched above or we can take effective constraints at
any stage of the stabilization algorithm to represent the constraint
surface.  This second method is the one applied in the geometrization of
Dirac algorithm \cite{Gotay}.  Geometrically, the relevant information is the
constraint surface and not the specific determination of the functions
one uses to describe it.  Both ways to realize the stabilization procedure
are equivalent.  The second is
advantageous from the algorithmic point of view, while the first is more
suitable if one wants to construct, for instance, the generators of the
gauge transformations \cite{GR}.

\section*{Acknowledgements}

We would like to thank Larry Shepley for a carefull 
reading of the manuscript.  J.M.P.
acknowledges support by the CICIT (contract numbers AEN95-0590 and
GRQ93-1047).
J.A.G. is supported by CONACyT postgraduate fellowship
and also thanks the Departament d'Estructura i Constituents de la
Mat\`eria at the Universitat de Barcelona for its hospitality.



\begin{thebibliography}{99}

\bibitem{FJ} L. Faddeev and R. Jackiw, Phys. Rev. Lett. 60 (1988)1692;
R. Jackiw, (Constrained) quantization without tears, in:
Proc. 2nd Workshop on constraints theory and quantization methods,
Montepulciano, 1993 (World Scientific, Singapore, 1995).

\bibitem{Dirac}
P.A.M Dirac, Can. J. Math. 2, 129 (1950); Proc. R. Soc.
London
Ser. A 246, 326 (1958); Lectures in Quantum Mechanics (Yeshiva University,
New
York, 1965).

\bibitem{GP} J. Antonio Garc\'\i a and Josep M. Pons, Equivalence of
Faddeev--Jackiw and Dirac approaches for gauge theories, {\it Int. J. Mod.
Phys. \bf A12} (1997) 451.


\bibitem{Siegel} W. Siegel, Conformal invariance of extended spinning particle
mechanics, {\it Int. J. Mod. Phys. \bf A3}, (1988) 2713.

\bibitem{GR} X. Gr\`acia and J. Roca, Covariant and noncovariant gauge
transformations for the conformal particle, {\it Mod. Phys. Lett \bf A8},
(1993) 1747.

\bibitem{W-alg} J. Gomis, J. Herrero, K. Kamimura and J. Roca,
Particle Mechanics Models with ${\cal W}$-symmetries, {\it
Ann. Phys. \bf 244} (1995) 67.

\bibitem{Klauder} John R. Klauder, Quantization of Systems with Constraints,
hep/th 9612025.

\bibitem{ghp} J. Gomis, M. Henneaux and J.M. Pons, Existence theorem
for gauge symmetries in Hamiltonian constrained dynamics, {\it
Class. Quamtum Grav. \bf 7} (1990) 1089.

\bibitem{Barcelona} C. Batlle, J. Gomis, J.M. Pons and N. Roman-Roy, J.
Math.
Phys. 27, 2953, (1986).


\bibitem{Henneaux} M. Henneaux and C. Teitelboim,
Quantization of gauge systems, (Princeton University Press, 1992).

\bibitem{cabo} A. Cabo and D. Louis-Martinez, On Dirac's conjecture
for Hamiltonian systems with first and second class constraints, {\it
Phys. Rev. \bf D42} (1990) 2726.

\bibitem{Gotay} M.J. Gotay, J.M. Nester and G. Hinds, J. Math. Phys. 19,
2388,
(1978); M.J. Gotay and J.M. Nester, Ann. Inst. H. Poincar\'e A 30 (1979) 129;
Ann. Inst. H. Poincar\'e A 32 (1980) 1.

\bibitem{GP2} J. Antonio Garc{\'i}a and J. M. Pons, Gauge generators and
Noether symmetries: A general approach. (In preparation).

\bibitem{BGGP} C. Batlle, J. Gomis, X. Gr\`acia and J. M. Pons, J. Math.
Phys. 30, 1345 (1989).

\bibitem{Zanelli} M. Henneaux, C. Teitelboim and J. Zanelli, Gauge
invariance and the gauge of freedom count, {\it Nucl. Phys. \bf B332}
(1990) 169.

\bibitem{Shepley} J.M. Pons and L.C. Shepley, {\it Class. Quantum
Grav. \bf 12} (1195) 1771.


\bibitem{BBH} G. Barnich, M. Henneaux and F.  Brandt, {\it
Commun. Math. Phys.} 174 (1995) 57.





\end{thebibliography}
\end{document}